\documentclass[12pt,a4paper]{article}
\usepackage{amsmath}
\usepackage{amssymb}
\usepackage{physics}
\usepackage{geometry}
\usepackage{slashed}  
\usepackage{braket}
\usepackage{graphicx}
\usepackage{xcolor}
\usepackage{tcolorbox}
\tcbuselibrary{skins}
\usepackage{bm}
\usepackage{authblk}
\usepackage{setspace}

\newcommand{\bea}{\begin{equation}\begin{aligned}}
\newcommand{\eea}{\end{aligned}\end{equation}}

\definecolor{myblue}{RGB}{0, 102, 204}
\usepackage[colorlinks=true, linkcolor=myblue, citecolor=myblue, urlcolor=myblue]{hyperref}

\begin{document}

\begin{titlepage}
\centering
\vspace*{3cm}

{\Large \bfseries Expanding Our Horizons: Rethinking Gravity's Role in the Quantum-to-Classical Transition \par}





\vspace{1.2cm}

{\normalsize \textbf{Aurora Ireland$^{1}$} \par}

\vspace{0.8cm}

August 24, 2026

\vspace{1.2cm}

{\small $^{1}$ Leinweber Institute for Theoretical Physics, Stanford University, Stanford, CA 94305, USA \par}

\vspace{0.5cm}

{\ttfamily anireland@stanford.edu}
\vspace{1.5cm}

\textbf{Abstract}
\vspace{0.5cm}

\begin{minipage}{0.9\textwidth}
\small 
The origin of cosmic structure is widely regarded as quantum, yet the Universe today appears classical. Standard lore attributes this to a ``quantum-to-classical'' transition on super-horizon scales during inflation. Gravity plays a central role: super-horizon dynamics squeeze quantum states, while the cosmological horizon enforces a system-environment split, leading to decoherence. But are these mechanisms always sufficient? We revisit this question, identifying assumptions and limitations in conventional arguments. We highlight recent work showing that beyond slow roll, non-linear dynamics of cosmological perturbations can generate and amplify quantum coherence at the closed-system level. Preliminary results further suggest that, for the minimal irreducible contribution to decoherence coming from stochastic kicks at the coarse-graining scale, the rate of coherence generation can outpace the decoherence rate. This raises the possibility that signatures of a quantum origin may persist in cosmic structure. We propose a phase-space analysis based on the Wigner function as a concrete route to identifying and probing such signatures. 
\end{minipage}

\vfill

{\small Essay written for the Gravity Research Foundation 2026 Awards for Essays on Gravitation}
\end{titlepage}

\section*{Introduction}

In the standard inflationary picture, the origin of cosmic structure is inherently quantum. Tiny fluctuations of the quantum vacuum, such as the metric perturbations $\mathcal{R}$ and $h_{ij}$, are stretched to super-horizon scales, where they ``freeze out''. Once inflation ends and these fluctuations re-enter the horizon, they become the density perturbations that seed the structures seen today. Their statistics should then be encoded in observables like the anisotropies of the Cosmic Microwave Background (CMB) and the large-scale structure of the Universe.

Despite their quantum origin, we routinely describe these fluctuations as classical stochastic fields. This is done explicitly in frameworks like stochastic inflation, but also implicitly in more general settings, like cosmological perturbation theory. This practice has traditionally been justified by invoking a certain ``quantum-to-classical'' transition said to occur while modes are outside the horizon.\footnote{To avoid directly addressing the quantum measurement problem, we adopt an operational definition of ``classical'': we will say that a system is classical if all of its statistics of interest can be reproduced by an appropriately constructed effective stochastic description.} There are two distinct arguments for why such a transition should occur. 

Even if perturbations are approximated as a closed quantum system evolving unitarily, super-horizon dynamics during slow roll drives the state into a highly squeezed configuration. This makes the non-commuting nature of the field and its conjugate momentum increasingly inaccessible to the late-time observables of interest, whose statistics then appear indistinguishable from those of a classical stochastic variable. This underlies the familiar ``decoherence without decoherence'' picture.

In reality, perturbations do not form a closed system. Long-wavelength modes interact and become entangled with shorter-wavelength modes we do not observe. Tracing over these environmental degrees of freedom produces decoherence, which suppresses coherences in the reduced state. The cosmological horizon provides a natural system-environment split, while the self-interactions of general relativity provide an irreducible coupling between the two. Thus, even if closed-system dynamics fail to render the state classical, one might expect environmental decoherence to finish the job.

These arguments suggest that gravitational dynamics alone should be sufficient to drive the quantum-to-classical transition, inevitably transforming the quantum fluctuations of inflation into effectively classical perturbations. However, this conclusion relies on assumptions which have received surprisingly little scrutiny. The closed-system argument relies on either linearized or slow-roll evolution, while the effectiveness of decoherence depends on a competition between the destruction of quantum coherence and dynamics that generate it. Once such effects are taken into account, is gravitational-driven classicalization really inevitable?

In this essay, we argue that it is not. In backgrounds beyond slow roll, non-linear interactions can generate quantum coherences, which are subsequently amplified by the same squeezing dynamics usually invoked to argue for classicality. Preliminary results indicate that, for physically motivated parameters, this generation of coherence can outpace its destruction by decoherence, allowing quantum effects to survive long enough to influence the statistics of late-time observables.
This opens up the possibility that the quantum origin of structure may leave observable relics in the Universe today.

\section*{Classicalization at the closed-system level}

\subsubsection*{Squeezing and the traditional lore}

There exists a long-standing argument that closed-system dynamics suffice to classicalize perturbations through so-called ``decoherence without decoherence''~\cite{Guth:1985ya,Polarski:1995,Lesgourgues:1996,Kiefer:1998qe}. The basic idea is that super-horizon evolution during inflation drives the quantum state into a highly squeezed configuration. In this limit, the perturbation and its conjugate momentum 
effectively collapse onto a single degree of freedom, rendering their commutator negligible compared with the classical correlations encoded by the anti-commutator. The system then ``looks'' classical, in the sense that its statistics are reproducible by a stochastic ensemble  for the late-time observables typically considered. 

To formalize this intuition, consider a Fourier mode of the curvature perturbation $\hat{\mathcal{R}}_k$ and its conjugate momentum $\hat{\pi}_k$. Mathematically, the closed-system argument states:
\bea
    \frac{[\hat{\mathcal{R}}_k, \hat{\pi}_k]}{\langle \{\hat{\mathcal{R}}_k, \hat{\pi}_k \} \rangle }  \sim \frac{\mathcal{D}_k}{\mathcal{G}_k} \sim e^{- 2 r_k} \ll 1 \,.
\eea
Parametric amplification outside the horizon drives $\langle \{\hat{\mathcal{R}}_k, \hat{\pi}_k \} \rangle \gg [\hat{\mathcal{R}}_k, \hat{\pi}_k]$. In the Heisenberg picture, this corresponds to a large hierarchy between the growing $\mathcal{G}_k$ and decaying $\mathcal{D}_k$ modes, while in the Schr{\" o}dinger picture it appears as a large squeezing parameter $r_k$. In all cases, the left-hand side decreases exponentially as the mode spends time outside the horizon. For modes relevant to the CMB, which remain outside for $\Delta N \sim 50$ $e$-folds, this suppression is enormous.

This argument has several limitations. Most importantly, it applies only at the linear level of the free theory, where states remain Gaussian in phase space. Beyond linear order, interactions generate non-Gaussianities, producing interference patterns in phase space that cannot simply be squeezed away. Moreover, squeezing itself is neither an intrinsic measure of classicality nor inherently meaningful; its amplitude can always be changed (or even eliminated) via a canonical transformation~\cite{Grain:2019vnq}, so it should not be expected to capture fundamental properties of the state. Finally, a squeezed state is still a highly quantum state; it just appears classical for the coarse-grained observables accessible in late-time cosmology. In fact, there is something rather paradoxical about using squeezing to argue for classicalization, when actually the same super-horizon dynamics behind squeezing are also responsible for producing strong entanglement between Fourier modes with opposite momenta, placing the CMB in a very quantum state~\cite{Martin:2015qta}. We thus need a sharper diagnostic of classicality.

\subsubsection*{Classicality and the Wigner function}

The Wigner function $W(x,p)$ provides a natural tool for this purpose. Formally, $W$ is defined as the Wigner-Weyl transform of the density operator $\hat{\rho}$,
\bea\label{eq:Wgendef}
    W(x, p) = \frac{1}{2\pi} \int_{-\infty}^{\infty} \dd u\, e^{- i p u} \left\langle x - \frac{u}{2} \big\vert \, \hat{\rho} \, \big\vert x + \frac{u}{2} \right\rangle \,,
\eea
where $(x,p)$ are coordinates on phase space~\cite{Wigner:1932}. Since this transformation is invertible, $W$ contains the same information as the density operator. The Wigner function can be used to formulate quantum mechanics on phase space, where it plays the role of a quasi-probability distribution. While $W$ is always real and normalized, it need not be positive. Only if $W$ is globally positive can it be interpreted as a true probability distribution. Negativity in $W$ reflects the presence of interference between different branches of the wavefunction, and so serves as an indicator of non-classicality~\cite{Kenfack:2004}.

Interference terms in $\hat{\rho}$ lead to oscillatory terms in $W$. For a simple illustration, consider the state $\ket{\psi} = \frac{1}{\sqrt{2}} \big( \ket{\psi_1} + \ket{\psi_2} \big)$, with $\psi_{1,2}$ Gaussian wavefunctions. The Wigner function can be written as
\bea\label{eq:W:cat}
    W = \frac{1}{2} (W_1 + W_2) + W_{\rm int} \,,
\eea
where $W_{1,2}$ are the Gaussian Wigner functions of $\psi_{1,2}$ and $W_{\rm int} \propto \cos(\Delta \bar{x} p + \Delta \bar{p} x)$, where $\Delta \bar{x} = \braket{\psi_1 | \hat{x} | \psi_1} - \braket{\psi_2 | \hat{x} | \psi_2}$ and $\Delta \bar{p} = \braket{\psi_1 | \hat{p} | \psi_1} - \braket{\psi_2 | \hat{p} | \psi_2}$. Notably, the interference term is oscillatory, taking both positive and negative values. 

This behavior is not unexpected. In fact, there is a theorem due to Hudson~\cite{Hudson:1979} which says that the only pure state with a globally positive Wigner function is a single Gaussian. Any deviation from this---even the superposition of two Gaussians---generically introduces negativity somewhere in phase space. Thus, going beyond linear theory and including interactions will necessarily produce negativity in $W$. 

From the perspective of reproducing classical-looking observables, what matters is not the presence of negativity, but whether it is operationally accessible. Expectation values are obtained by integrating the Wigner function against the phase-space representation of observables,
\bea
    \langle \mathcal{O} \rangle_{\rm stoch} = \int dx dp \, W(x,p) \mathcal{O}(x,p) \,.
\eea
If the relevant observables do not have support where $W<0$, then these quantum features do not contribute, and the system will appear classical. 

The question, then, is whether gravitational dynamics during inflation inevitably suppress Wigner negativity, or whether it can persist in a way which leaves imprints on cosmological observables.


\subsubsection*{When perturbations refuse to classicalize}

This question can be addressed by computing the Wigner function for the curvature perturbation $W(\mathcal{R}, \pi_\mathcal{R})$ in generic inflationary backgrounds, as was recently done in Ref.~\cite{Ireland:2026txt}.\footnote{The computation in Ref.~\cite{Ireland:2026txt} was technically performed using the Goldstone of spontaneously broken time translations in the effective field theory (EFT) of inflation~\cite{Cheung:2007}. The Goldstone coincides with $\mathcal{R}$ (up to a constant prefactor) at the linear level, but is also well-defined non-perturbatively.} In particular, this work focused on constant-roll backgrounds~\cite{Motohashi:2014}, which are defined by the condition that the inflaton's acceleration and friction terms remain in fixed proportion throughout evolution,
\bea
    \ddot{\phi} = \beta H \dot{\phi} \,, \,\,\, \text{with} \,\,\, \beta = \text{constant} \,.
\eea
Such evolution is well-motivated and reduces to familiar cases like slow roll for $\beta \approx 0$ and ultra-slow roll for $\beta \approx -3$. Backgrounds with $\beta \neq 0$ exhibit so-called ``non-attractor'' dynamics; the inflaton's velocity $\dot{\phi}$ is not uniquely determined by its position $\phi$, so the system does not reduce to single-clock evolution. As a result, the curvature perturbation continues evolving on super-horizon scales rather than freezing, with important consequences for the dynamics and interactions of perturbations.

In particular, during non-attractor phases, key quantities encoding the background dynamics, like the Hubble-flow parameter $\epsilon_1 = - \dot{H}/H^2$, evolve non-trivially in time. In a sense which can be formalized in the EFT of inflation~\cite{Cheung:2007}, the effective couplings of perturbations inherit this time dependence, generating interaction terms directly from the background evolution. Such interactions should be distinguished from the gravitational self-interactions of perturbations, which are always present but typically suppressed by powers of $\epsilon_1 \ll 1$. In other words, during non-attractor phases the background evolution itself acts as a source of non-linearity and non-Gaussianity.

These interactions impact the structure of the quantum state in phase space, and hence the Wigner function. While linear evolution (corresponding to a quadratic Hamiltonian) preserves Gaussianity in $W$, non-linear evolution from interaction terms produces a departure from Gaussianity, oscillations, and regions of negativity. The degree of negativity in the Wigner function therefore serves as a diagnostic of whether these interactions are important on super-horizon scales.

\begin{figure}[t!]
\centering
\includegraphics[width=0.9\textwidth]{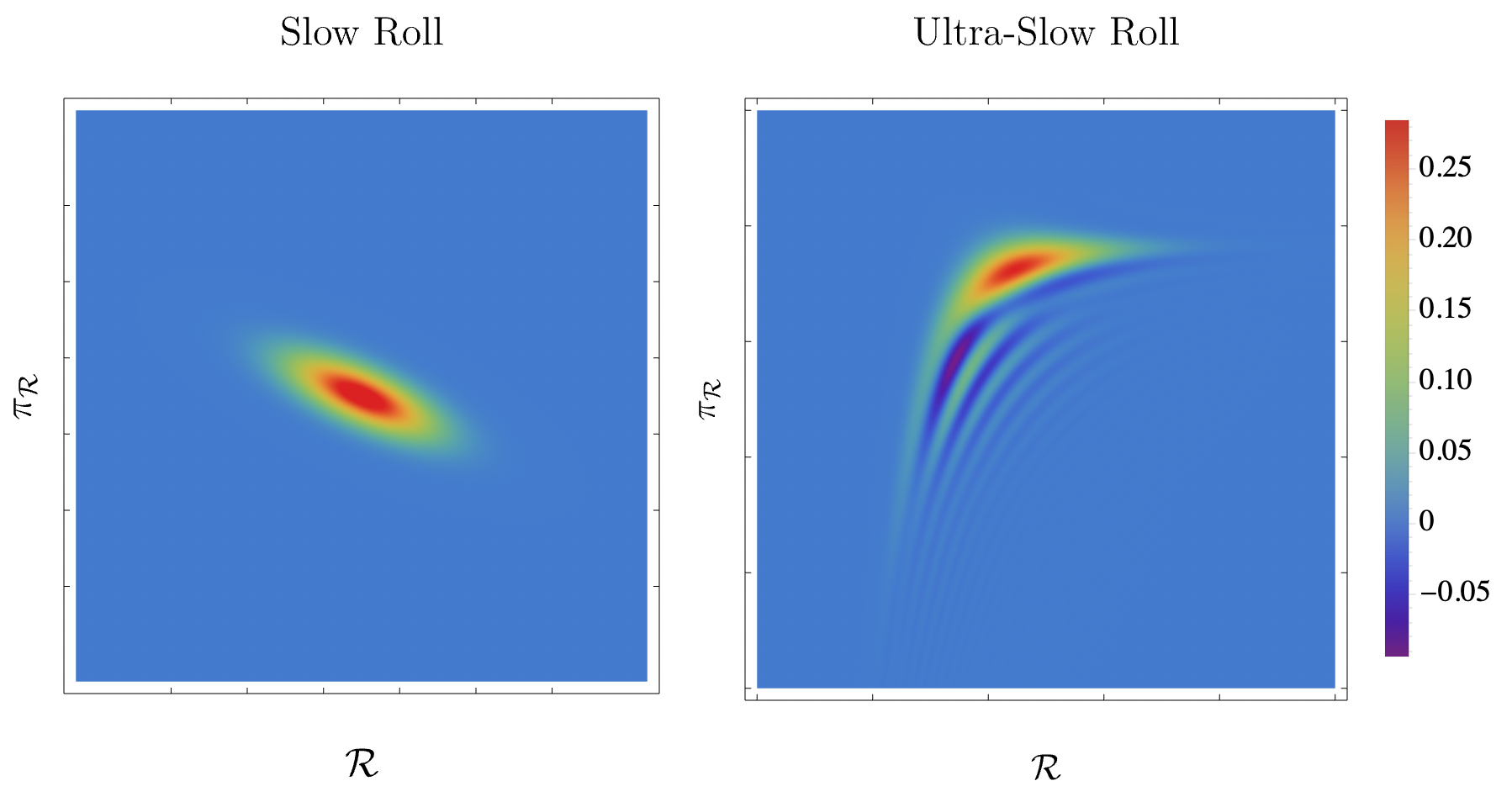}
\caption{Schematic density plot of the Wigner function $W(\mathcal{R}, \pi_\mathcal{R})$ in slow roll (left) and ultra-slow roll (right) inflationary backgrounds. The former is a globally positive Gaussian while the latter displays interference fringes and regions of negativity.}
\vspace{0.5\baselineskip}
\label{fig:Wigner}
\end{figure}

This intuition is reflected in the behavior shown in Fig.~\ref{fig:Wigner}. In slow roll, and more generally in any background analyzed at the linear level, one recovers the usual result that $W$ is globally positive, with elliptical contours in phase space that get squeezed with increasing time (see left panel). By contrast, in non-attractor backgrounds ($\beta \neq 0$), the Wigner function develops pronounced interference fringes, with an oscillatory structure radiating outwards in phase space that includes regions of negativity (see right panel). This Wigner negativity reveals the presence of genuine quantum coherences, reflecting the non-classical nature of the state.

These results are not limited to constant-roll models, but are expected in any non-attractor phase. Because the Wigner contours are no longer elliptical, squeezing fails to align the phase-space coordinates along a single direction, undermining the ``decoherence without decoherence'' mechanism described above. 

Notably, the amount of Wigner negativity, as quantified by the negativity volume 
\bea
    \mathcal{N} = \frac{1}{2} \bigg( \int \dd \mathcal{R} \dd \pi_\mathcal{R} \, |W(\mathcal{R},\pi_\mathcal{R})| - 1 \bigg) \,,
\eea
increases with time on super-horizon scales. For example, in ultra-slow roll backgrounds it grows exponentially with the number of $e$-folds as $\mathcal{N} \propto e^{2 N}$~\cite{Ireland:2026txt}. This occurs despite the simultaneous increase in squeezing, consistent with our earlier assertion that squeezing does not imply classicality. In fact, squeezing amplifies non-linearities, which in turn increases Wigner negativity.

These results indicate that quantum effects, as diagnosed by Wigner negativity, can survive and even grow on super-horizon scales during non-attractor periods of inflation, contrary to conventional wisdom. Accordingly, gravitational dynamics do not inevitably classicalize the state at the closed-system level. This raises the intriguing possibility that signatures of the quantum origin of cosmological perturbations may be more accessible than previously thought, a possibility we will return to in later sections.

\section*{Classicalization at the open-system level}

\subsubsection*{Cosmological spacetimes are open systems}

Although cosmological perturbations in the early Universe are often approximated as a collection of decoupled Fourier modes evolving independently, this picture is incomplete; in reality, they should be modeled as an open system. The reason for this is rooted in the causal structure imposed by the cosmological horizon along with the non-linear nature of general relativity.

Present-day observables, such as the CMB anisotropies and large-scale structure, originate from long-wavelength perturbations which were super-horizon during inflation. By contrast, shorter-wavelength modes that were sub-horizon oscillate rapidly and largely average out, rendering them unobservable. Since we only measure the long-wavelength sector, it defines our \textit{system}, while the unobserved short-wavelength modes act as an \textit{environment} for it. This intuition underlies the stochastic inflation formalism~\cite{Starobinsky:1986fx}, in which the field is split into short- and long-wavelength components relative to a coarse-graining scale $k_\sigma = \sigma a H$ set by the horizon size, and the system is described by effective open-system dynamics.

The cosmological horizon thus provides a natural decomposition into system and environment. Moreover, the non-linear nature of general relativity guarantees a minimal set of interactions coupling system with environment, even in the absence of additional fields~\cite{Burgess:2022nwu}. As a result, the long-wavelength sector must be treated as open. 


\subsubsection*{Decoherence and Wigner negativity}

The open-system nature of cosmological perturbations inevitably introduces decoherence~\cite{Joos:1984}. Historically, this has been quantified using measures like the purity or entanglement entropy. Like squeezing, however, these do not always admit a straightforward interpretation in terms of classicality. For example, situations can arise where the purity decays even while the quantum discord increases or Bell inequalities become more strongly violated~\cite{Martin:2021znx}. Moreover, these quantities depend sensitively on the chosen coarse-graining and do not map directly onto observables. By contrast, the Wigner function provides a more faithful diagnostic, with a direct connection to observables.

For the reduced density matrix $\hat{\rho}_{\rm red} = \text{Tr}_{\rm env}[ \hat{\rho}]$ obtained by tracing out the environmental degrees of freedom, decoherence manifests as a suppression of the off-diagonal elements in the basis selected by the system-environment interaction (the ``pointer basis''~\cite{Zurek:1981}). The resulting system then behaves like a classical statistical mixture rather than a coherent superposition. In phase space, decoherence acts to erase interference features in the Wigner function, decreasing Wigner negativity. 

Consider again the Wigner function of Eq.~(\ref{eq:W:cat}). Decoherence suppresses the off-diagonal density matrix elements, such that in the completely-decohered limit, the density operator reduces to $\hat{\rho}_{\rm dec} \approx \frac{1}{2} ( \ket{\psi_1}\!\bra{\psi_1} + \ket{\psi_2}\!\bra{\psi_2} )$. In this limit, the interference term then vanishes $W_{\rm int} \approx 0$, and $W \approx \frac{1}{2} (W_1 + W_2)$ reduces to the sum of two Gaussians---which is everywhere positive.

Because most interactions, including gravitational self-interactions, couple locally in field space, the pointer basis is approximately the field basis $\ket{\mathcal{R}}$. Decoherence therefore predominantly suppresses coherence between different field amplitudes, damping oscillations in the conjugate momentum direction. How quickly and efficiently this occurs, however, depends on the details of the interactions.

\subsubsection*{Non-attractor backgrounds}

Most existing decoherence estimates assume slow-roll backgrounds or rely on linear perturbation theory. One might expect that going beyond these settings to non-attractor phases could qualitatively change the picture. In particular, one might naively worry that the enhanced interactions in these backgrounds could strengthen the coupling with the environment, enhancing the decoherence rate, just as they amplify Wigner negativity. It is therefore not immediately obvious which effect wins.

This question can now be addressed directly. Follow-up work to Ref.~\cite{Ireland:2026txt} extends the analysis to incorporate open-system effects like decoherence using the open EFT of inflation~\cite{Salcedo:2024smn}, which supplements the unitary action with local operators encoding noise and dissipation. Preliminary results confirm that these terms behave as expected, erasing interference fringes and reducing Wigner negativity at a rate controlled by the EFT coefficient. Fixing this coefficient to the natural value corresponding to stochastic kicks imparted by the short-wavelength modes as they cross the coarse-graining scale, we find that the rate of negativity growth exceeds the rate of erasure due to decoherence. That is, there is a net generation of quantum coherence even after these open system effects are taken into account. 

In order to make sense of this result, keep in mind that long- and short-wavelength modes couple to the background spacetime in fundamentally different ways. Only the former experience a parametric amplification outside the horizon; for the latter, the gradient term $\propto k^2$ dominates such that they remain approximately oscillatory regardless of the background dynamics. We should therefore not expect their background-induced interactions to be enhanced in the same way as those of the long-wavelength sector. Instead, they remain suppressed by the usual factors of the Hubble flow parameter $\epsilon_1$ and powers of the inverse Planck mass.

While we stress that these results are preliminary and account only for the minimal irreducible contribution to decoherence guaranteed by gravity, they nevertheless indicate that incorporating open-system effects need not restore classicality. This raises the exciting possibility that quantum coherences generated during inflation may survive long enough to leave observable signatures in the statistics of late-time observables.

\section*{Observational prospects}

In this essay, we have examined whether gravitational dynamics during inflation inevitably drive primordial perturbations toward apparent classicality. We revisited the standard arguments for classicalization at the closed-system level, highlighting assumptions and shortcomings, and showed that once interactions are included, non-attractor dynamics can instead generate and amplify quantum coherence. We then considered the effect of open-system effects like decoherence. Preliminary results indicate that, for the minimal irreducible contribution expected from gravitational interactions coming from short-wavelength modes of the same field, the rate at which quantum coherence grows outpaces the decoherence rate, resulting in net coherence generation for the relevant modes. These findings suggest that gravitational dynamics need not erase the quantum character of primordial perturbations before it can influence their late-time statistics. 

Establishing whether cosmic structure is truly of quantum mechanical origin would represent a profound advance in our understanding of the early Universe. Identifying observables that can answer this question unambiguously is notoriously challenging. Proposals based on violations of cosmic Bell inequalities are either extremely difficult to realize~\cite{Martin:2017zxs} or require elaborate, non-minimal inflationary scenarios unlikely to be realized in nature~\cite{Maldacena:2015bha}. Another approach is to search for features in the primordial bispectrum or higher-point correlation functions~\cite{Green:2020whw}, though such signatures can be model-dependent or experimentally challenging to detect. While there is in principle a large amount of quantum information encoded in the CMB, it exists largely in Fourier space~\cite{Martin:2015qta}, and does not readily translate to real-space observables.

Despite these challenges, the program initiated in Ref.~\cite{Ireland:2026txt} provides reason for optimism and points to a practical route towards identifying observable signatures of quantum coherences. In particular, the phase space structure of the Wigner function offers a concrete guide for anticipating where such signatures might appear. For example, if oscillatory features are confined to the tails of the distribution, their first observable consequences are likely to arise in the rare events which probe these tails, rather than in lower-point statistics like the power spectrum. Moreover, by examining how the Wigner-Weyl transform of an operator overlaps with regions of negativity, one can design targeted observables capable of directly probing these quantum features. Developing such observables and determining their detectability in realistic cosmological data would provide a path toward putting our understanding of the quantum-to-classical transition to observational test.

\section*{Acknowledgements}
I thank Thomas Colas and Vincent Vennin for illuminating conversations and helpful suggestions. I also thank Dan Hooper and Marisa Von Drasek for feedback on an earlier version of this essay. I acknowledge support from NSF Grant PHY-2310429, Simons Investigator Award No.~824870, DOE HEP QuantISED award \#100495, the Gordon and Betty Moore Foundation Grant GBMF7946, and the U.S.~Department of Energy (DOE), Office of Science, National Quantum Information Science Research Centers, Superconducting Quantum Materials and Systems Center (SQMS) under contract No.~DEAC02-07CH11359.

\bibliographystyle{JHEP}
\bibliography{biblio}

\end{document}